\documentclass[aip,apl,superscriptaddress]{revtex4-1}
\usepackage{amsfonts}
\usepackage{amsmath}
\usepackage{amssymb}
\usepackage{bm}
\usepackage{graphicx}
\usepackage[T1]{fontenc}
\begin{document}

\title{Optimized nonlinear terahertz response of
graphene in a parallel-plate waveguide} 

\author{Parvin Navaeipour}
\affiliation{Department of Physics, Engineering Physics and Astronomy,Queen's University, Kingston, Ontario K7L 3N6, Canada}

\author{Marc M. Dignam}
\affiliation{Department of Physics, Engineering Physics and Astronomy,Queen's University, Kingston, Ontario K7L 3N6, Canada}

\email{p.navaeipour@queensu.ca}

\begin{abstract}
Third harmonic generation of terahertz radiation is expected to occur in monolayer graphene due to the nonlinear relationship between the crystal momentum and the current density. In this work, we calculate the terahertz nonlinear response of graphene inside a parallel-plate waveguide including pump depletion, self-phase, and cross-phase modulation. To overcome the phase mismatching between the pump field and third-harmonic field at high input fields due to self-phase and cross-phase modulation, we design a waveguide with two dielectric layers with different indices of refraction. We find that, by tuning the relative thicknesses of the two layers, we are able to improve phase matching, and thereby increase the power efficiency of the system by more than a factor of two at high powers. With this approach, we find that dispite the loss in this system, for an incident frequency of $2$ THz, we are able to achieve power efficiencies of $75 \%$ for graphene with low Fermi energies of $20$ meV and up to $35\%$ when the Fermi energy is $100$ meV.
\end{abstract}

\maketitle

\section{Introduction}
Graphene, as a zero-bandgap two-dimensional semiconductor with a linear electron band dispersion near the Dirac points has the potential to exhibit very interesting nonlinear optical properties \cite{mikailov,Mikhailov2010,Gu,Glazov}. The linear dispersion relation of the electrons near the Dirac points leads to a constant electron speed \cite{castro,sarma}. Thus, the intraband current induced in graphene by terahertz (THz) fields displays clipping as the amplitude of the incident field increases, which generates odd harmonics in the current and transmitted electric field \cite{Bowlan2014,Alnaib2015,Ibra2015,ibraheem2014}. Exploiting the nonlinear response of graphene enables one to produce higher frequency THz radiation through the generation of harmonics. Several experimental and theoretical
groups have examined third-harmonic generation from graphene at terahertz frequencies. Almost all have employed a configuration where the field is normally
incident on the graphene \cite{Hafez,Paul,PBowlan}. However, here we consider a configuration where the radiation propagates in a metallic parallel-plate waveguide (PPW), with the graphene sheet lying at the midpoint between the two plates as shown in Fig.~1 \cite{parvin}.
\begin{figure}
  \centering
  \includegraphics[width=0.8 \columnwidth]{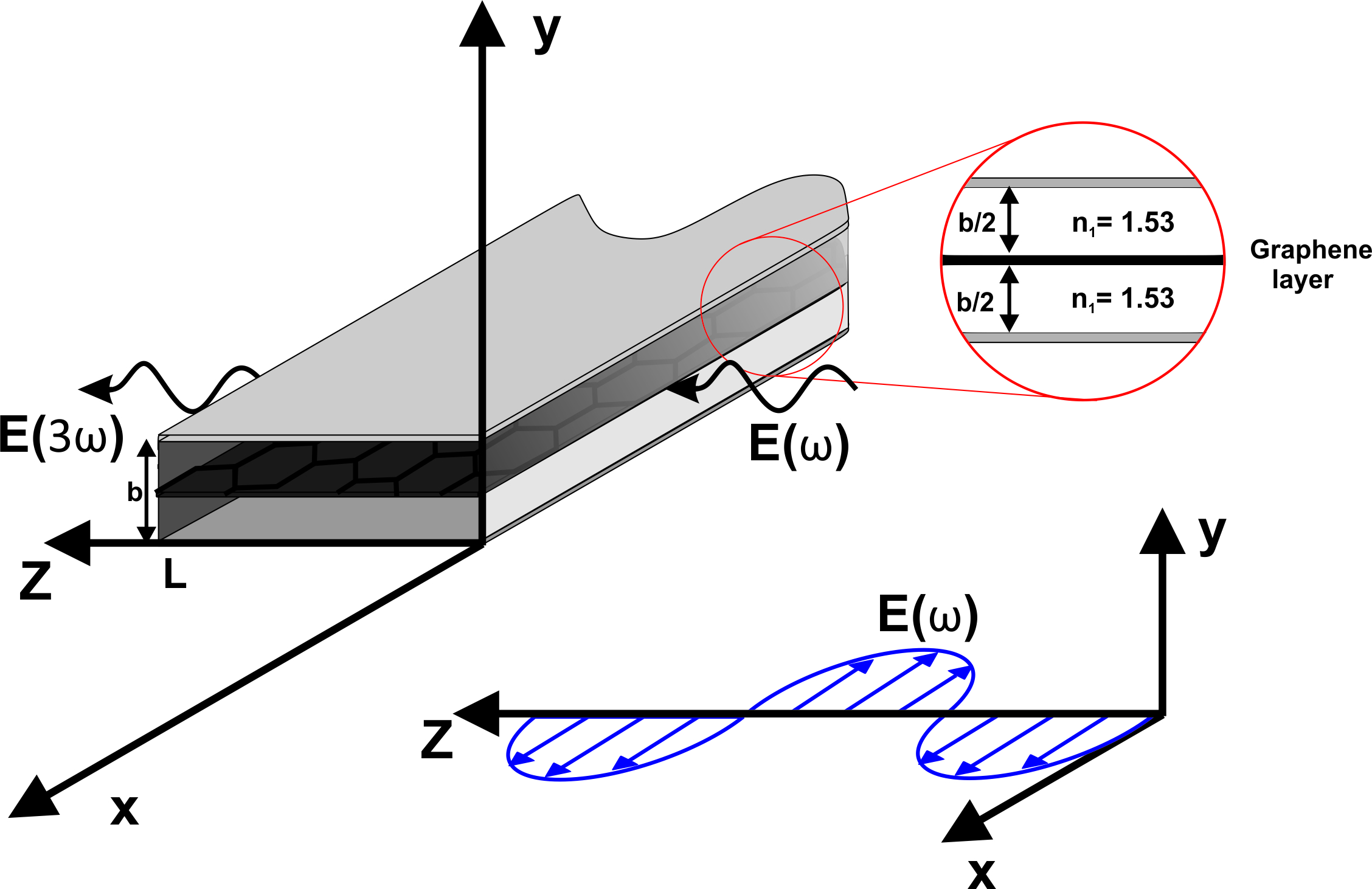} 
  \caption{The metallic parallel-plate waveguide with graphene inside which forms the system being modelled. The inner material of the waveguide is polyolefin and the graphene is placed at the center of the waveguide at $ y = b/2$. The pump field propagates in the $+z$-direction and is polarized in the $x$-direction.}
  \label{fig:Figure1}
\end{figure}
With this configuration, we increase the interaction time between the radiation and graphene, and thereby generate a larger-amplitude harmonic field. We have shown in previous work that this configuration can increase the power efficiency of the system by more than a factor of $100$, relative to the results for the normal-incidence configuration and that the power efficiency is relatively insensitive to the plate separation, but depends strongly on the Fermi energy \cite{parvin}. However, in that work, we did not include the effects of pump depletion, self-phase modulation (SFM), and cross-phase modulation (XFM) in our calculations. In this work, we develop a coupled-mode theory including all propagating lossy modes to calculate the power efficiency for third-harmonic generation in a PPW and  use this model to examine the impact of these effects on the power conversion efficiency. \\
For weak input fields, there is generally good phase matching in the waveguide between the pump field in the TE$_1$ mode at $\omega$ and third harmonic field in the TE$_3$ mode at $3\omega$. However, as we shall show, when the pump field amplitude increases, the phase matching degrades due to SFM and XFM. To overcome this, we propose a new configuration in which the waveguide contains two layers of dielectric materials: cyclic polyolefin ($n_1= 1.53$) and phenol formaldehyde resin ($n_2 = 1.70$) \cite{Nestor}. One goal in this work is to optimize the thickness of the dielectric layers and the Fermi energy of the graphene to obtain phase matching and thereby maximize the generated third-harmonic electric field.\\ 
The paper is organized as follows. In Sec.~\textrm{II} we expand the electric field at the fundamental and third harmonic in terms of the lossy modes of the PPW and use the slowly-varying envelope approximation to derive the differential equations for the mode amplitudes. In Sec.~\textrm{III} we compare the results obtained for the generated third-harmonic field, using our coupled-mode theory in the undepleted pump approximation, with pump depletion, and using a full calculation, which includes pump depletion and self- and cross-phase modulation. In Sec.~\textrm{IV} we propose new configuration PPW and demonstrate that this configuration allows us to essentially eliminate phase mismatch over a wide range of Fermi energies and input field amplitudes. Finally, in Sec.~\textrm{V} we summarize our results.  
\section{theory}
In this section, we first solve for the lossy linear modes of the waveguide with graphene present. We then expand the fields at $\omega$ and $3 \omega$ in terms of these linear modes, to derive the nonlinear coupled mode equations.\\
Our parallel-plate waveguide consists of two metallic plates placed at y = 0 and y = b, with the graphene midway between the plates at $y = b/2$ as shown in Fig.~1. The inner material
of the waveguide is chosen to be cyclic polyolefin, with a refractive index of $n_1 = 1.53$, due to its compatibility with graphene, ease of fabrication and low loss at THz frequencies~\cite{Nestor}. The THz wave propagates in the $+z$ direction and is polarized in the $x$ direction. For simplicity we take the plates to be perfect conductors that are infinite in the $x$ direction.\\ 
From Maxwell's equations, we obtain the inhomogeneous wave equation,
\begin{equation}
\nabla \times \nabla \times \mathbf{\mathcal{E}(\mathbf{r}, t)} = \mu_0 \epsilon_0 \dfrac{\partial^2 \mathbf{\mathcal{E}(\mathbf{r}, t)}}{\partial t^2} + \mu_0 \dfrac{\partial \mathbf{\mathcal{J}}(\mathbf{r},t)}{\partial t},
\end{equation}
where $\mathbf{\mathcal{E}(\mathbf{r}, t)}$ is the total electric field and $\mathbf{\mathcal{J}}(\mathbf{r}, t)$ is the total current density in the graphene. The incident electric field is taken to be harmonic with frequency of $\omega$ and a third-harmonic electric field is generated, so that the total electric field is 
\begin{align}
\mathbf{\mathcal{E}(\mathbf{r}, t)} =& \, \mathbf{E}(\mathbf{r};\omega)e^{-i\omega t} + c.c. \\
+& \, \mathbf{E}(\mathbf{r};3\omega)e^{-i3\omega t} + c.c. \nonumber
\end{align}
and the total current density is
\begin{align}
\mathbf{\mathcal{J}}(\mathbf{r}, t) =& \,  \mathbf{J}(\mathbf{r}; \omega) e^{-i\omega t} + c.c.\\
+& \, \mathbf{J}(\mathbf{r}; 3\omega) e^{-i3\omega t} + c.c.. \nonumber
\end{align}
Now, the current density can be broken up into its linear and nonlinear components as $\mathbf{J} = \mathbf{J}_L  + \mathbf{J}_{NL} $. Here, $\mathbf{J}_L$ is the current due to the linear conductivity of the graphene and is given by 
\begin{equation}
\mathbf{J}_L (\mathbf{r}; \omega^{\prime}) = \sigma^{(1)} ( \omega^{\prime}) \mathbf{E}(\mathbf{r}; \omega^{\prime}) \delta(y- b/2),
\end{equation}
where $\sigma^{(1)} ( \omega^{\prime}) $ is the linear conductivity of the graphene, with $\omega^{\prime} = \lbrace \omega, 3\omega \rbrace$. The linear conductivity has both intraband and interband contributions \cite{Alnaib2015,parvin,horng}. At low Fermi energies ($E_F \leq 20$ meV), a photon with frequency of $f_0 \equiv \omega_0/2\pi = 2$ THz is not able to cause interband transitions. Thus, the dominant contribution to the linear conductivity is the intraband conductivity at $\omega_0$ \cite{parvin}. However, for the third harmonic, the interband transition does contribute to some degree in the linear conductivity at low Fermi energies and so is included in our calculations \cite{parvin}. The intraband contribution to the linear conductivity at zero temperature is proportional to the Fermi energy and is given by 
\begin{equation}
\sigma^{(1)}(\omega) = \dfrac{e^2 \tau E_F}{\pi \hbar^2 (1 +i\omega \tau)},
\end{equation}
where $E_F$ is the Fermi energy and $\tau$ is phenomenological scattering time, which in this work is taken to be $50 \, fs$. Thus, to limit linear loss, it is better to work at low Fermi energies.\\ 
The nonlinear current density of the graphene, $\mathbf{J}_{NL}$, arises from the third-order nonlinear conductivity, $ \sigma^{(3)} $. At $3\omega$ it is given by 
\begin{align}
\mathbf{J}_{NL}(\mathbf{r}; 3\omega) =& \, \mu_0 \sigma^{(3)} (3\omega; \omega, \omega, \omega) \Big\lbrace \mathbf{E} (\mathbf{r};\omega) \Big\rbrace ^3 \delta(y-b/2)\\
+& \, 3\mu_0 \sigma^{(3)} (3\omega; 3\omega, -3\omega, 3\omega)  \mathbf{E}(\mathbf{r};3\omega)  \vert \mathbf{E} (\mathbf{r};3\omega) \vert^2 \delta(y-b/2) \nonumber\\
+& \, 6\mu_0 \sigma^{(3)} (3\omega; 3\omega, -\omega, \omega)  \mathbf{E}(\mathbf{r};3\omega)  \vert \mathbf{E} (\mathbf{r};\omega) \vert^2 \delta(y-b/2). \nonumber
\end{align}
The first term in Eq.~(6) is the most important, as it is the source of the third-harmonic electric field. The next two terms are respectively 
related to SFM and XFM.\\
The nonlinear current density at $\omega$ is given by,
\begin{align}
\mathbf{J}_{NL}(\mathbf{r};\omega) =& \,  3\mu_0 \sigma^{(3)} (\omega; 3\omega, -\omega, -\omega)  \mathbf{E}(\mathbf{r};3\omega) \Big\lbrace \mathbf{E}^{\ast} (\mathbf{r};\omega) \Big\rbrace^2 \delta(y-b/2) \\
+& \, 3\mu_0 \sigma^{(3)} (\omega; \omega, -\omega, \omega) \mathbf{E}(\mathbf{r};\omega) \vert \mathbf{E}(\mathbf{r};\omega) \vert ^2 \delta(y-b/2)  \nonumber \\
+& \, 6\mu_0 \sigma^{(3)} (\omega; \omega, -3\omega, 3\omega)  \mathbf{E}(\mathbf{r};\omega) \vert  \mathbf{E} (\mathbf{r};3\omega) \vert^2 \delta(y-b/2). \nonumber
\end{align}
The first term in Eq.~(7) gives the nonlinear current at the graphene that results in to pump depletion and the next two terms represent SPM and XFM, respectively. The factors of $3$ and $6$ in Eqs.~(6) and (7) arise from the number of ways of generating the nonlinear polarization in those cases.\\
In Eqs.~(6) and (7), $\sigma^{(3)} $ is the nonlinear conductivity of the graphene. In this work, we use the theoretical expression of Cheng \emph{et al}. \cite{cheng}, which is derived from perturbative calculations at zero temperature, for electrons close to the Dirac points, with the neglect of the effects of the scattering ($\tau \rightarrow \infty$). Under these assumptions,
\begin{align}
\sigma^{(3)} (3\omega; \omega, \omega, \omega) = \dfrac{i \sigma_0(\hbar v_F e)^2}{48 \pi (\hbar \omega)^4} T(\dfrac{\hbar \omega}{2 E_F}),
\end{align}
where $\sigma_0 = \dfrac{e^2}{4\hbar}$ is the universal conductivity of the graphene, $v_F$ is the Fermi velocity of the electrons in the graphene, taken to be $1.1 \times 10^6$ m/s and
\begin{equation}
T(x) \equiv 17 G(x) - 64 G(2x) + 45G(3x),
\end{equation} 
in which, 
\begin{equation}
G(x) \equiv \ln(\dfrac{1 + x}{1 - x})+i \pi \, \Theta(\vert x \vert -1), 
\end{equation}
where $ \Theta (x)$ is the Heaviside step function. The other nonlinear conductivities in Eqs.~(6) and (7) are related to $\sigma^{(3)} (3\omega; \omega, \omega, \omega)$ by
\begin{align}
\sigma^{(3)} (3\omega; \omega, \omega, \omega) =&  \, \sigma^{(3)} (3\omega; 3\omega, -3\omega, 3\omega) = \sigma^{(3)} (3\omega; 3\omega, -\omega, \omega) \\ 
=& \, 3\sigma^{(3)} (\omega; 3\omega, -\omega, -\omega) = 3\sigma^{(3)} (\omega; \omega, -\omega, \omega) = 3\sigma^{(3)} (\omega; \omega, -3\omega, 3\omega). \nonumber 
\end{align}
\subsection{Linear Modes} 
For harmonic waves travelling in the $+z$ direction with
angular frequency $\omega$, the linear electric field for the $n^{th}$ transverse electric (TE) mode is given by
\begin{align}
E_x^{(n)} (y,z;\omega) =
\Bigg\lbrace
  \begin{tabular}{ccc}
  $E_ne^{i\tilde{\beta}_n(\omega)z}\sin[\tilde{k}_n(\omega)y]$  \quad \quad \quad \quad \quad $y>b/2 $  \\
  $- E_n e^{i\tilde{\beta}_n(\omega)z}\sin[\tilde{k}_n(\omega)(y-b/2)]$  \quad $y<b/2 $
  \end{tabular}
  \end{align}
where $ n = 1, 2, 3, ... $, $E_n$ is the amplitude of the n$^{th}$ mode, and $\tilde{k}_n$ is the complex wave number for the field's $y$ dependence. This wavenumber depends on the linear conductivity of the graphene and is obtained by enforcing the boundary conditions at the graphene, which leads to the following transcendental equation \cite{parvin} 
\begin{equation}
\sigma^{(1)}(\omega) = \dfrac{4i\tilde{\phi}_n}{\omega \mu b}\cot(\tilde{\phi}_n),
\end{equation}
where $\tilde{\phi}_n \equiv \dfrac{\tilde{k}_n(\omega) b}{2}$ and $\mu$ is the permeability of the dielectric. The complex propagation constant, $\tilde{\beta}_n$, of the TE$_{n}$
mode is given by
\begin{equation}
\tilde{\beta}_n(\omega) =\sqrt{\left(\dfrac{n_1 \omega}{c_0} \right)^2 - \tilde{k}_n^2},
\end{equation}
where $c_0$ is the speed of light in vacuum and $n_1$ is the refractive index of the dielectric material. If there is no graphene, i.e., for a \emph{bare waveguide}, $\tilde{k}_n \equiv k_n^0 = n\pi/b$, where $n$ is
an integer.\\
\subsection{Coupled Mode Equations}
In this work, we take the input field at $\omega$ to be in the TE$_{1}$ mode at $z = \, 0$. We expand the field at $\omega$ in terms of the lossy TE${_n}$ modes as
\begin{align}
{E}(y,z;\omega) = \sum_{n} A_n(z;\omega) e^{i\tilde{\beta}_n(\omega)z} \sin(\tilde{k}_n (\omega) y)
\end{align}
and expand the generated third-harmonic electric field as 
\begin{align}
{E}(y,z;3\omega) = \sum_{n} A_n(z;3\omega) e^{i\tilde{\beta}_n(3\omega)z} \sin(\tilde{k}_n (3\omega) y), 
\end{align}
where the summation is over all of the lossy modes propagating in the waveguide and the $A_n(z;\omega)$ are slowly varying envelopes. Although this is not an exact expansion (as the lossy modes are not complete), we showed in our previous work \cite{parvin} that using this expansion in the undepleted pump approximation, we obtain almost identical results for the generated third-harmonic field as were obtained using an exact Green function approach as long as the frequency is not close to the cut-off frequency. \\
The initial conditions are: $A_n(0;\omega) = \delta_{n,1}{E_{input}}/{\sin(\tilde{k}_n(\omega)\dfrac{b}{2})} $ and $A_n(0;3\omega) = 0$ where $E_{input}$ is the amplitude of the incident field at the graphene. We now employ our mode expansions to solve Eq.~(1) for the fundamental and third harmonic fields including pump-depletion, SFM, and XFM. Using Eqs.~(6), (7) and (14) in Eq.~(1) along with the facts that $\nabla \cdot \mathbf{E} = \, 0$ and the modes are essentially orthogonal gives
\begin{align}
- \nabla^2 {E}(y,z;3\omega) =& \, i3\omega \mu_0 \sigma^{(1)}(3\omega) {E}(y,z;3\omega)\delta(y-b/2) \\
 +& \, 3i \omega \mu_0 \sigma^{(3)}(3\omega; \omega, \omega,\omega) \Big\lbrace {E}(y,z;\omega) \Big\rbrace^3  \delta(y-b/2)\nonumber\\
 +& \, 3i\omega \mu_0 \sigma^{(3)} (3\omega; 3\omega, -3\omega, 3\omega)  {E}(y,z;3\omega) \vert {E} (y,z;3\omega) \vert^2 \delta(y-b/2) \nonumber \\
 +&  \, 6i \omega \mu_0 \sigma^{(3)} (3\omega; 3\omega, -\omega, \omega) {E}(y,z;3\omega) \vert {E}(y,z;\omega) \vert ^2  \delta(y-b/2) \nonumber \\
 +& \, 9\mu_0 \epsilon \omega^2 {E} (y,z;3\omega). \nonumber
\end{align}
Now, using Eqs.~(15) and (16) and employing the slowly-varying envelope approximation (i.e neglecting $\dfrac{d^2 A_n(z;3\omega)}{dz^2}$) we obtain the following differential equation for the amplitude of the electric field at $3\omega$ for the $m^{th}$ mode (See appendix for details):
\begin{align}
\dfrac{d A_m  (z;3\omega)}{dz}   & =  \dfrac{-3 \omega \mu_0}{b \tilde{{\beta}}_m (3\omega)}  \bigg\lbrace \sum_{n^{\prime} }   \sigma^{(3)} (3 \omega; \omega, \omega,\omega) \Big\lbrace A_{n^{\prime}}  (z;\omega) \Big\rbrace^3 \nonumber \\
& \; \; \; \; \; \; \; \; \times  e^{i(3 {\tilde{\beta}}_{n^{\prime}}(\omega)- \tilde{{\beta}}_m (3\omega))z}   S_m^{ \ast (3)} \Big( \dfrac{b}{2}\Big) \Big\lbrace S_{n^{\prime}}^{(1)}\Big(\dfrac{b}{2}\Big) \Big\rbrace ^3  \\
 &    + \sum_{n }\sigma^{(3)} (3 \omega; 3\omega, -3\omega,3\omega)   A_n (z; 3\omega) \vert  A_{n}  (z;3\omega)\vert ^2 \nonumber \\
 & \; \; \;  \; \; \; \; \; \times e^{i( {2\tilde{\beta}}_{n}(3\omega) - \tilde{\beta}_{m}(3\omega) - {\tilde{\beta}}_{n}^{\ast} (3\omega))z} 
  S_m^{ \ast (3)}\Big( \dfrac{b}{2}\Big)  S_{n}^{\ast (3)}\Big( \dfrac{b}{2}\Big)  \Big\lbrace S_{n}^{ (3)}\Big( \dfrac{b}{2}\Big)\Big\rbrace^2 \nonumber \\
   &  + \sum_{n }  \sum_{n^{\prime}} 2\sigma^{(3)} (3 \omega; 3\omega, -\omega,\omega)   A_n (z; 3\omega) \vert  A_{n^{\prime}}  (z;\omega)\vert ^2 \nonumber \\
 & \; \; \;  \; \; \; \; \; \times e^{i( {\tilde{\beta}}_{n}(3\omega) - \tilde{\beta}_{m}(3\omega) + \tilde{\beta}_{n^{\prime}}(\omega) - {\tilde{\beta}}_{n^{\prime}}^{\ast} (\omega))z} 
  S_m^{ \ast (3)} \Big( \dfrac{b}{2}\Big) S_{n^{\prime}}^{\ast (1)} \Big( \dfrac{b}{2}\Big) S_{n^{\prime}}^{ (1)}\Big(\dfrac{b}{2}\Big)  S_n^{ (3)}\Big( \dfrac{b}{2}\Big) \bigg\rbrace, \nonumber
\end{align}
where, $S_n^{(1)} \Big( y \Big) \equiv \, sin(\tilde{k}_n(\omega)y)$ and $S_n^{(3)} \Big(y \Big) \equiv \, sin(\tilde{k}_n(3\omega)y)$. Similarly, for the electric field modes at $\omega$ we obtain:  
\begin{align}
\dfrac{d A_m (z;\omega)}{d z} =&   \dfrac{-3 \omega \mu_0}{b \tilde{\beta}_m (\omega)}  \sum_{n} \bigg\lbrace  \sum_{ n^{\prime}}  \sigma^{(3)} (\omega; 3\omega,-\omega,-\omega) A_{n^{\prime}} (z;3\omega) \Big\lbrace A_n^{\ast } (z;\omega) \Big\rbrace^2 \nonumber \\
&  \; \; \; \; \; \; \; \; \; \; \; \; \; \; \; \; \; \;  \; \; \; \; \; \; \times e ^{i(\tilde{{\beta}}_{n^{\prime}}(3\omega) - 2\tilde{\beta}^{\ast}_n(\omega)- \tilde{\beta}_m (\omega))z}  {S_m^{\ast}}^{(1)} \Big(\dfrac{b}{2} \Big) S_{n^{\prime}}^{(3)} \Big(\dfrac{b}{2} \Big)  \Big\lbrace S_n^{\ast (1)} \Big(\dfrac{b}{2} \Big) \Big\rbrace^2\\
 & \; \; \; \; \; \; \; \; \; \; \; \; \; \; \;  \; \; \; \; \; \; + \sigma^{(3)} (\omega; \omega, -\omega,\omega) A_n(z;\omega) \vert A_n(z;\omega) \vert^2 \nonumber \\
&   \; \; \; \;\; \; \; \;\; \; \; \; \; \; \; \; \; \; \; \; \; \; \; \; \times e^{i(2\tilde{\beta}_n(\omega)- \tilde{\beta}_n^{\ast}(\omega) - \tilde{\beta}_m (\omega)) z}S_m^{\ast (1)} \Big(\dfrac{b}{2} \Big) \Big\lbrace S_n^{(1)} \Big(\dfrac{b}{2} \Big) \Big\rbrace^2  S_n^{\ast (1)} \Big(\dfrac{b}{2}\Big) \nonumber \\
&  \; \; \; \; \; \; \; \; \; \; \; \; \; \; \;  \; \; \; \; \; \; + \sum_{n^{\prime}} 2\sigma^{(3)} (\omega; \omega, -3\omega,3\omega) A_n(z;\omega) \Big\vert A_{n^{\prime}}(z;3\omega) \Big\vert^2 \nonumber \\
& \; \; \; \;\; \; \; \;\; \; \; \; \; \; \; \; \; \; \; \; \; \; \; \; \times e^{i(\tilde{\beta}_n(\omega)+ \tilde{\beta}_{n^{\prime}}(3\omega) -\tilde{\beta}_{n^{\prime}}^{\ast}(3\omega)- \tilde{\beta}_m (\omega)) z}S_m^{\ast (1)} \Big(\dfrac{b}{2}\Big) S_n^{(1)} \Big(\dfrac{b}{2}\Big)  \Big\vert S_{n^{\prime}}^{ (3)} \Big(\dfrac{b}{2}\Big) \Big\vert^2 \bigg\rbrace. \nonumber  
\end{align} 
\section{results}
In this section, we solve the coupled dynamic equations for the amplitudes of the electric fields at $3\omega$ and $\omega$, given by Eqs.~(18) and (19) respectively. In all that follows, we take the incident (pump) field to have frequency $f_0\equiv\omega_0/2 \pi = 2$ THz and take the plate separation to be $70$ $\mu m$. We choose this plate separation because it is the largest separation for which there are only two propagating modes in the waveguide at $3\omega$. Note also that for this plate separation, only the TE$_1$ mode is a propagating mode at $\omega$. In our previous work we showed that there is a perfect phase matching between the first mode at $\omega$ and third mode at $3\omega$ when there is no graphene. Thus, we only include first and third modes in our calculations. To solve the coupled equations of Eq.~(18) and (19), we employ a Runge-Kutta algorithm; solving these coupled equations numerically takes less than one minute on an i7 processor. \\
\begin{figure}[h]
  \centering
  \includegraphics[width=1 \columnwidth]{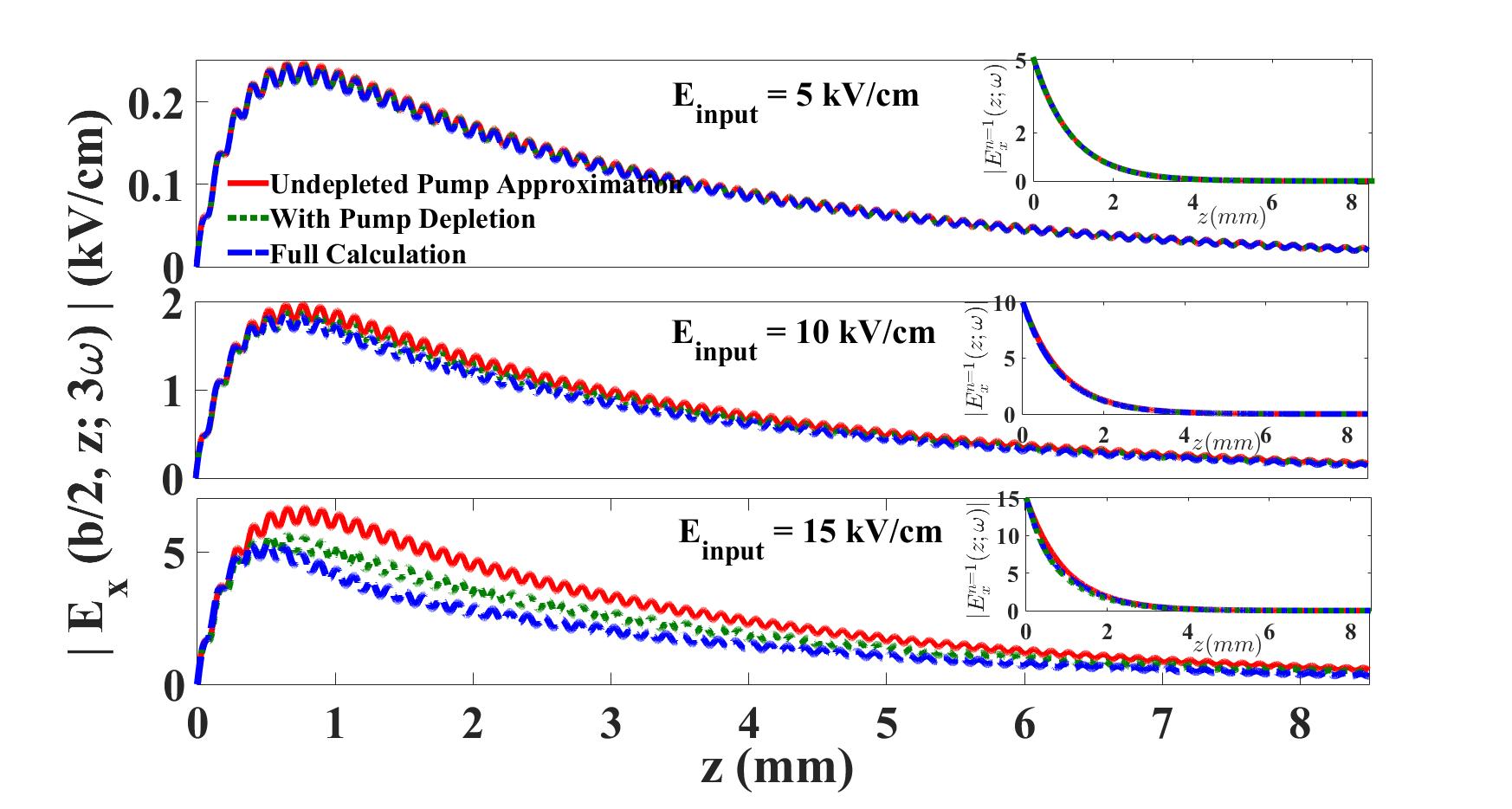} 
  \caption{Generated third-harmonic electric field at the graphene calculated in the undepleted pump approximation (red solid curves), calculated with pump depletion (green dotted curves) and calculated with full calculation (blue dashed curves) for different input fields of (a) $5$ kV/cm, (b) $10$ kV/cm and $15$ kV/cm for a Fermi energy of $E_F = 50$ meV. In the inset, we plot the fundamental field at the graphene as a function of $z$.}
  \label{fig:Figure3}
\end{figure} 
In Fig.~2, we plot the generated third-harmonic electric field at the graphene as a function of~$z$ for a Fermi energy of $E_F = 50$ meV for three different input field amplitudes ( $E_{input} = 5, 10$, and $ 15 $ kV/cm). We present the results of three different calculations: undepleted pump approximation (red solid curves), calculation with pump depletion (green dotted curves), and the full calculation that includes the SFM and XFM and pump depletion (blue dashed curves). To demonstrate the importance of pump depletion, we
first present the results of the calculation when the SFM, XFM and pump depletion are neglected. This is accomplished by keeping only the first term on the right hand side of Eqs.~(18) and (19) and taking $A_1(z;\omega) = A_1(0;\omega)$. Note however that we still include linear loss in all modes. Initially the third-harmonic grows rapidly, while the field at $\omega$ (see insets) decays exponentially until it is essentially gone after a propagation distance of about $2$ mm. Oscillations in the third-harmonic arise due to the phase mismatch between the third-harmonic field in the $n = \, 1$ and $n =\, 3$ modes, which is why they persist even after the fundamental field is essentially gone. Almost identical results are obtained for the generated third-harmonic field at low input fields when pump depletion is included, as seen in Fig.~2(a). However, as we increase the input field, we see a significant reduction in the generated third-harmonic field with pump depletion relative to the undepleted pump approximation. This is due to the increased transfer of power from the incident (pump) field to the third harmonic field. This can also be seen in the insets of Fig.~2(b) and (c), for strong input fields, where the pump field decays faster when the pump field is increased.\\
We now turn to the effects of self- and cross-phase modulation. As we increase the input field, we see in Fig.~2(b) and (c) that including SFM and XFM results in a decrease in the generated third-harmonic electric field. This is due to a degradation in the phase matching between the third mode at $3\omega$ and first mode at $\omega$ due to SFM and XFM. For an input field of $15$ kV/cm, this results in a  $6 \%$ reduction in the peak field. As we shall see, the effect is much more significant at lower Fermi energies and/or higher input fields. Let us now examine the effects of SFM and XFM on phase matching in the PPW in more detail. \\
It is easy to show, using Eq.~(14) that for a \emph{bare waveguide} there is a perfect phase matching between the $ n = 3$ mode at $3\omega$ and the first mode at $\omega$ \cite{parvin}. To generate a strong third-harmonic electric field we need to have a very small effective refractive index difference between these two modes in the presence of graphene loss and SFM and XFM. This effective refractive index difference for a lossy waveguide with SFM and XFM is approximately given by 
\begin{equation}
\Delta n_{eff} \equiv  n_{eff}^{(3)}(3\omega) - n_{eff}^{(1)}(\omega),
\end{equation}
where 
\begin{align}
n_{eff}^{(1)} (\omega) \equiv \dfrac{\text{Re}\lbrace \tilde{\beta}_1 (\omega) \rbrace}{\omega/c_0} + n_2^{(1)}(\omega) \vert E_{input} \vert^2 
\end{align}
and
\begin{align}
n_{eff}^{(3)} (3\omega) \equiv \dfrac{\text{Re}\lbrace \tilde{\beta}_3 (3\omega) \rbrace}{3\omega/c_0} + n_2^{(3)}(3\omega) \vert  E_{input} \vert^2. 
\end{align}
The first terms in each of Eqs.~(21) and (22) are the input-field-independent effective refractive indices for the first mode at $\omega$ and third mode at $3\omega$, respectively. The second terms are added to approximately account for the change in the effective index due to the SFM and XFM. Using Eqs.~(18) and (19) we obtain
\begin{align}
n_2^{(1)}(\omega) \simeq  \, & {\text{Im} \lbrace -\dfrac{3 \mu_0 c_0}{b \tilde{\beta}_1(\omega)} \sigma^{(3)}(\omega; \omega, -\omega, \omega)  \vert S_1^{(1)}\Big(\dfrac{b}{2}\Big) \vert^2 \rbrace},  \\
n_2^{(3)}(3\omega) \simeq  \, & {  \text{Im} \lbrace -\dfrac{ 2\mu_0 c_0}{b \tilde{\beta}_3(3\omega)} \sigma^{(3)}(3\omega; 3\omega, -\omega, \omega)  \vert S_3^{(3)}\Big(\dfrac{b}{2}\Big) \vert^2  \rbrace}. \nonumber
\end{align}
In deriving these expressions, we have only included the terms proportional to the square of the electric field at $\omega$ and have taken the pump field to be given by its value at $z = 0$. We find that the terms proportional to the square of the third-harmonic electric field are negligible relative to the linear electric field. However, in our full numerical calculations, all terms are retained, as is the $z$-dependence of $A_1(\omega)$.\\
The effective index difference between the third mode at $3\omega$ and the first mode at $\omega$ as a function of Fermi energy is shown in Fig.~3 for different input pump field amplitudes. When $E_{input} = 0$, $\Delta n_{eff} $ linearly increases with Fermi energy, due to the dependence of the propagation constants on the doping level of the graphene. As the input field increases, $\Delta {n_{eff}}$ increases for all Fermi energies, but increases the most for low Fermi energies. At the lowest Fermi energy of $20$ meV and input field of $15$ kV/cm, $\Delta n_{eff}$ is very large, reaching a value of approximately $0.06$. This strong dependence on Fermi energy has its origin in the strong dependence of the nonlinear conductivity on Fermi energy, as seen in Eqs.~(8) to (10).
\begin{figure}[h]
  \centering
  \includegraphics[width=0.8 \columnwidth]{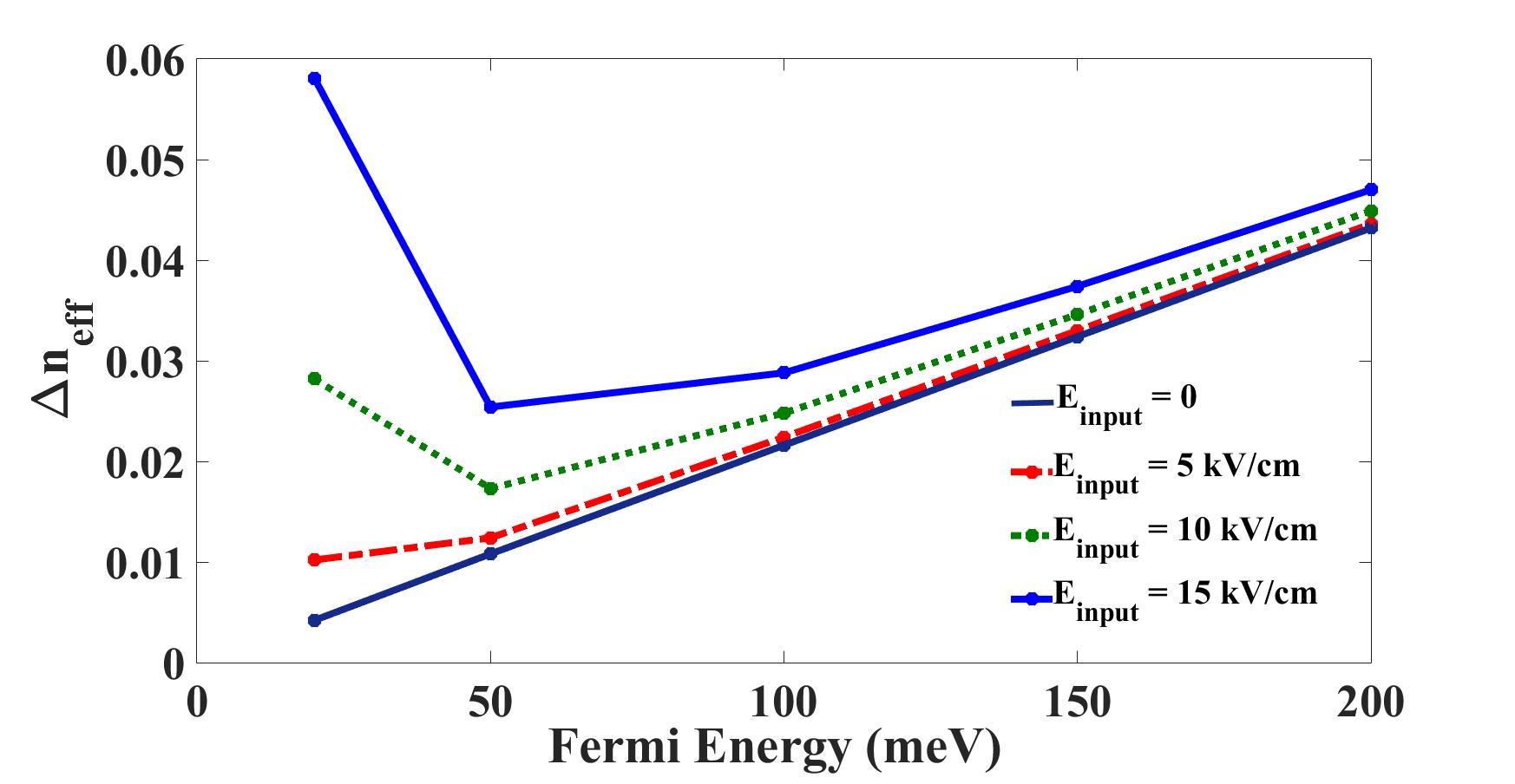} 
  \caption{Power-dependent effective refractive index difference between the TE$_3$ mode at $3\omega$ and the TE$_1$ mode at $\omega$, as a function of Fermi energy for four different input fields. When $E_{input}=0$, there are no effects due to XFM and SFM present. }
  \label{fig:Figure4}
\end{figure}\\
We now consider the power efficiency $S_{ eff}$ of the device, which is defined as the ratio of the power in the third harmonic at the end of the waveguide to the power in the fundamental at the beginning of the waveguide. In Fig.~4 we plot the maximum power efficiency as a function of Fermi energy in three different schemes: undepleted-pump approximation, with pump depletion, and full calculation. In all cases, the length of the waveguide is chosen to be the distance at which the power in the third harmonic field is a  maximum, as seen in Fig.~2. Decreasing the Fermi energy leads to higher nonlinear conductivity and lower linear conductivity by the graphene. Therefore, we obtain a higher power efficiency as the Fermi energy is decreased.   
\begin{figure}[h]
  \centering
  \includegraphics[width=0.8 \columnwidth]{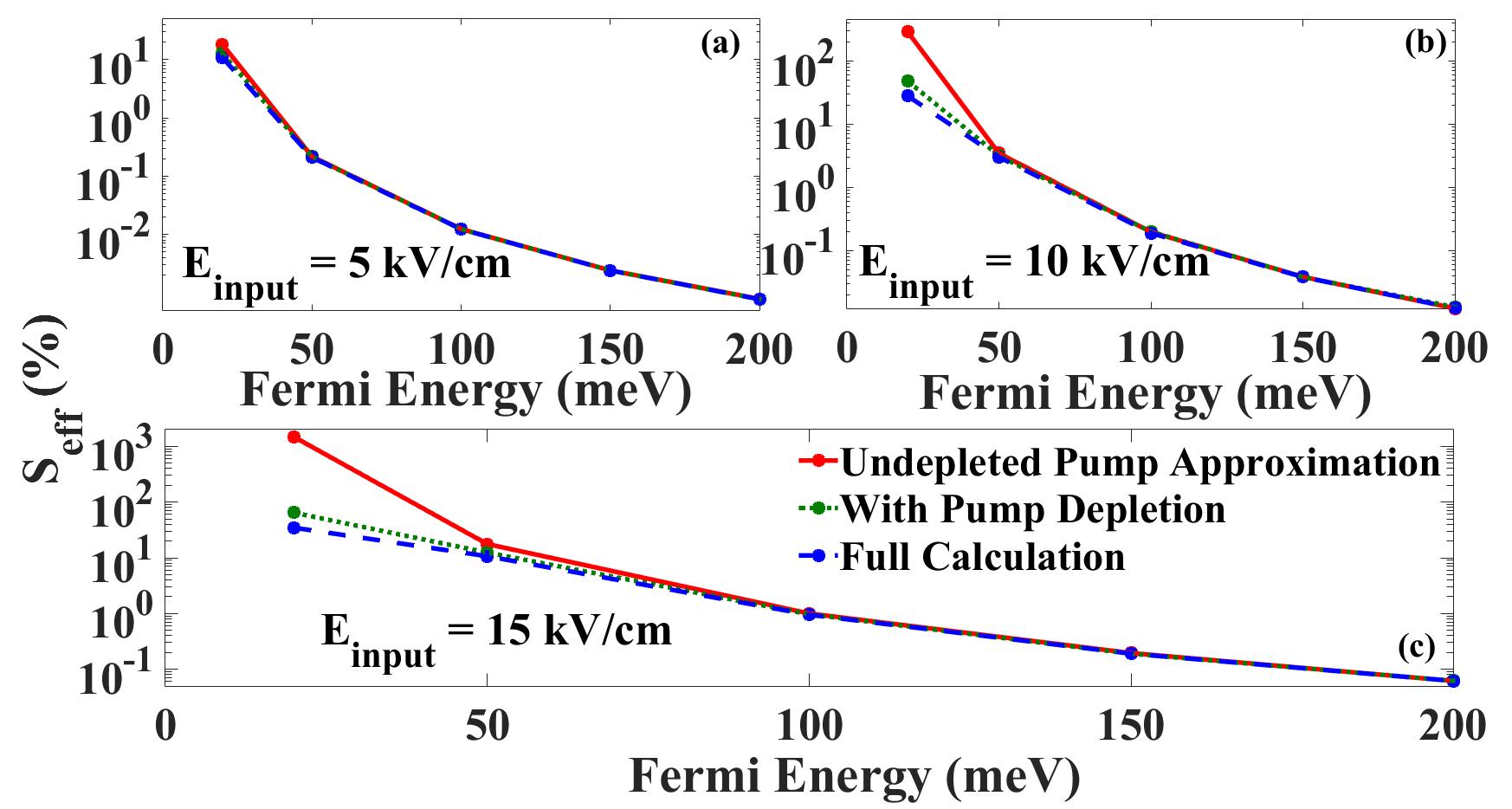} 
  \caption{Power efficiency of the waveguide as a function of Fermi energy calculated in the undepleted pump approximation (red solid line), calculated with pump depletion (green dotted line) and for the full calculation (blue dashed line) for different input fields of (a) $5$ kV/cm, (b) $10$ kV/cm and (c) $15$ kV/cm.} 
  \label{fig:Figure5}
\end{figure}
We see in Fig.~4(a) that for a weak input field of $5$ kV/cm, the power efficiency is almost the same in all three calculation schemes. It is only at low Fermi energies ($\sim 20$ meV) that there is a noticeable difference in the power efficiency. As we increase the input field, we see in Figs.~4(b) and (c) that the effects of pump depletion, SFM, and XFM become very significant, particularly at low Fermi energies. For strong enough fields and low Fermi energies, our calculations for the undepleted pump approximation yield a non-physical power efficiency that is greater than $100 \%$ \cite{parvin}. However, including pump depletion results in power efficiencies less than $100 \, \%$, as required. It is seen that the power efficiency decreases when SFM and XFM are included. For example, for Fermi energy of $E_F = 20$ meV, when the input field is $E_{input} = 10$ kV/cm, the power efficiency decreases from $S_{eff} = 48.26 \, \%$ in the PPW with pump depletion to $S_{eff} = 28.13 \, \%$ in the PPW with full calculation, while for an input field of $E_{input} = 15 $ kV/cm, the power efficiency decreases from $ 64.47 \, \%$ to $ 34.64 \, \%$. It is therefore worth examining if we can modify the structure to decrease the phase mismatch introduced by SFM and XFM. 
\section{New Configuration: Mitigating self- and cross-phase modulation}
In this section we define a new configuration of the PPW in order to deal with phase mismatching due to the SFM and XFM. We consider the waveguide shown in Fig.~5, where there are two different dielectric materials in the waveguide: cyclic polyolefin with refractive index of $n_1= 1.53$ and phenol-formaldehyde resin with refractive index of $n_2 = 1.70$. The graphene layer is located at $y = b/2$ midway between two metallic plates. The $n_1$ material is in the regions $y=0$ to $y=d_1$ and $y=b-d_1$ to $y=b$, while the $n_2$ material in the region $y=d_1$ to $y=b-b_1$. This new configuration allows us to control to some degree the phase matching between the third mode at $3\omega$ and the first mode at $\omega$. In the following, we optimize $d_1$ to obtain the best phase matching between the TE$_3$ mode at $3\omega$ and TE$_1$ mode at $\omega$, and thereby maximize the third harmonic generation and the power efficiency. Note that in all cases, the total plate separation is fixed at $b=70\, \mu m$.
\begin{figure}[h]
  \centering
  \includegraphics[width=0.7 \columnwidth]{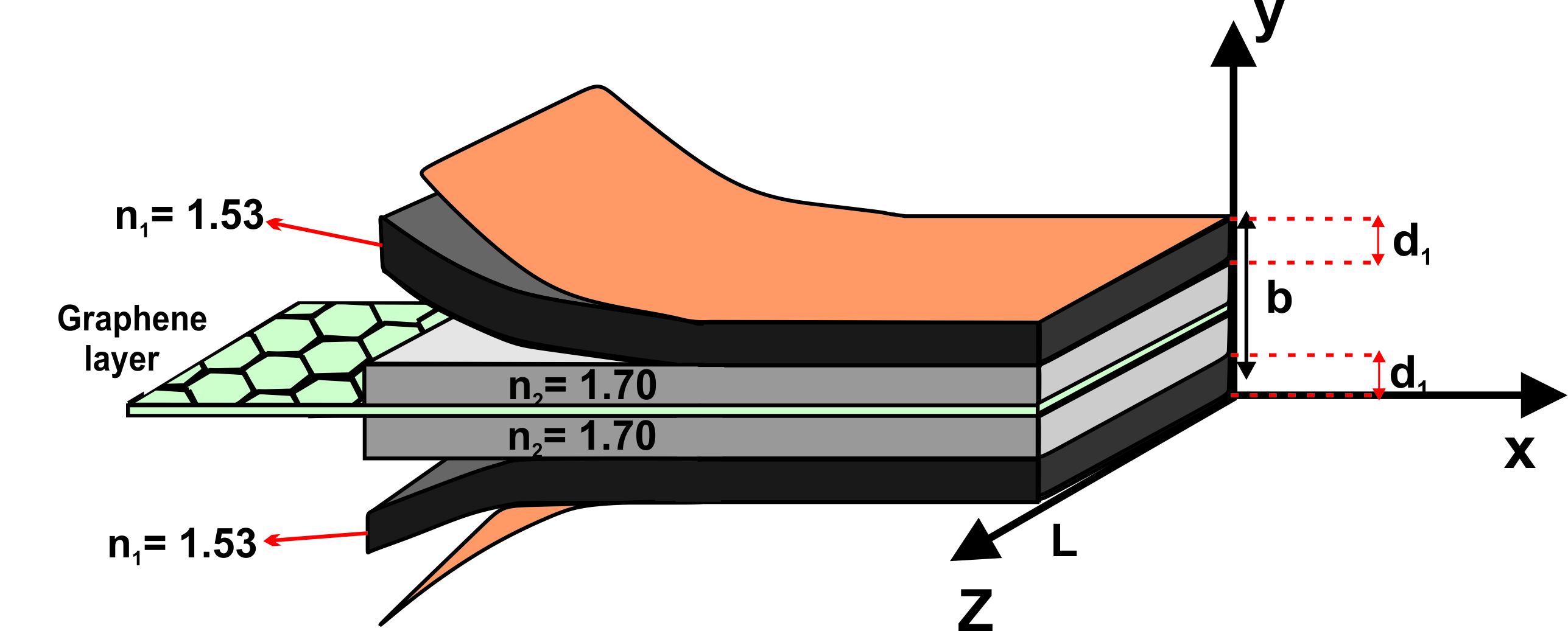} 
  \caption{New configuration parallel plate waveguide where we have two different dielectric layers with different indices of refraction. We use this structure to reduce phase mismatch by optimizing the thickness, $d_1$, of the material with the lower index.}
  \label{fig:Figure6}
\end{figure}\par  
\subsection{Phase Matching}
In this section we examine the effect of the layers thickness on the linear phase mismatch. We begin by examining the linear modes. In Fig.~6 we plot the normalized TE$_1$ mode at $\omega$ and TE$_3$ mode at $3\omega$ for a $2$ THz field in a $70$ $\mu m$ waveguide in our original configuration and in our new configuration. As an example we choose $d_1 = 25 \, \mu m$. Note that the TE$_1$ mode peaks at the centre of the waveguide, inside of the low-index ($n_1$) material, while the TE$_3$ mode also has peaks inside of the high-index ($n_2$) material. As a result, it is expected that by increasing the width of the high-index material, we can raise the effective index of the TE$_1$ mode more than that of the TE$_3$ mode and thereby modify the index mismatch. This new configuration can be used to not only help to decrease the phase mismatching induced by the SFM and XFM but to overcome the linear phase mismatch introduced by the graphene. Note also that the new configuration leads to an increase in the amplitude of the field at the graphene in the TE$_1$ mode. This will yield a slight increase in the generated field for a given input power.
\begin{figure}[h]
  \centering
  \includegraphics[width=0.7 \columnwidth]{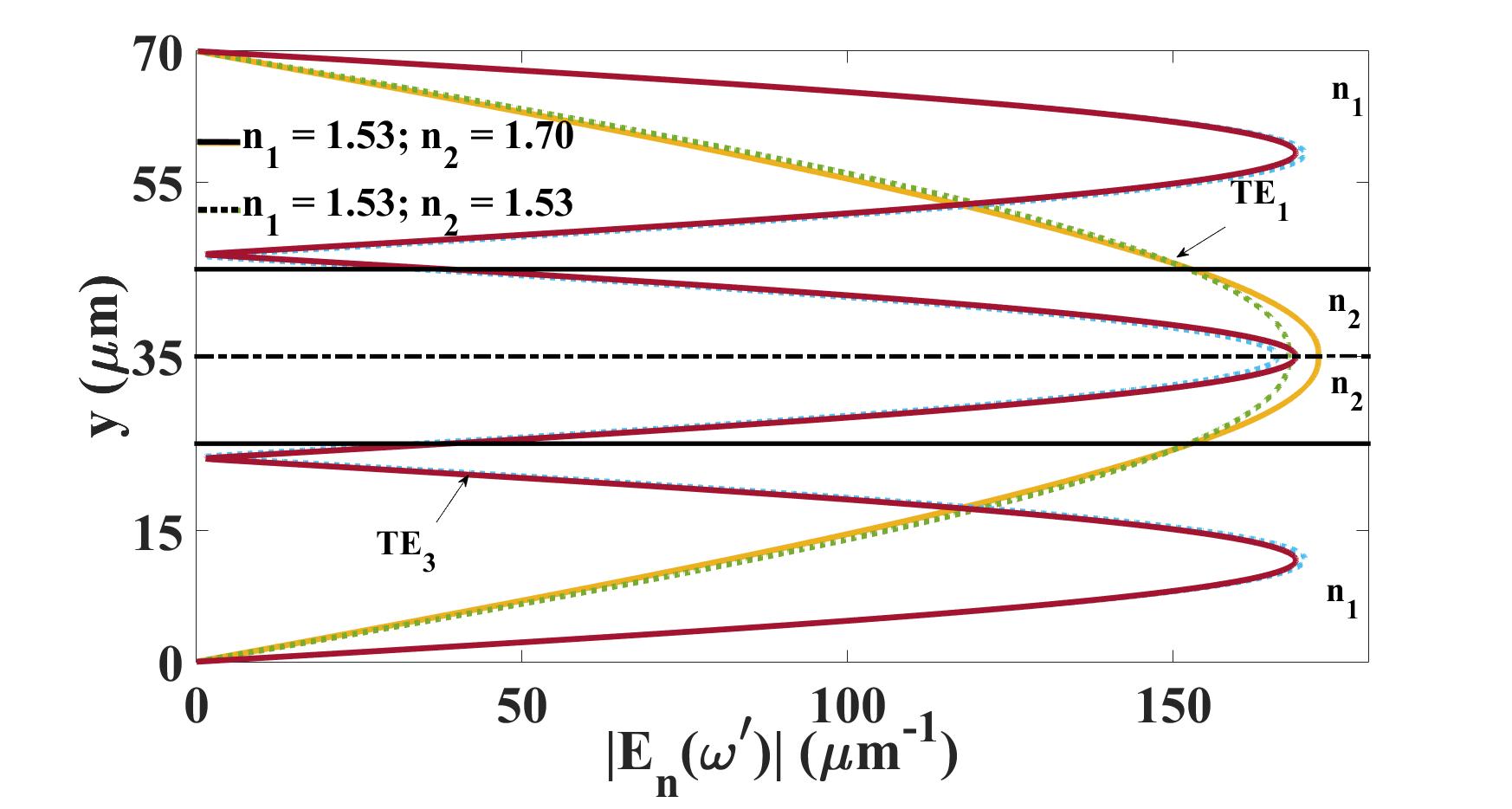} 
  \caption{ Absolute value of the electric field for the normalized TE$_1$ mode at $\omega$ and TE$_3$ mode at $3\omega$ at $z = 0$ for a $2$ THz field in the new configuration ( $n_1 = 1.53$ and $ n_2 = 1.70$) and original configuration ($n_1 = 1.53$ and $ n_2 = 1.53$) for $d_1 = 25$ $\mu m$ and $E_{F} = 50$ meV.}  
  \label{fig:Figure7}
\end{figure}\\
We now examine how the effective refractive index changes with the thickness of the region with lower refractive index ($d_1$). In Fig.~7, we plot the linear effective refractive index difference, $\Delta n_{eff}$ for a waveguide with plate separation of $b = 70 \, \mu m$ for different Fermi energies as a function of $d_1$. In this calculation, we set $n_2^{(1)} (\omega)$ and $n_2^{(3)} (3\omega)$ to zero. It is seen that perfect phase matching ($\Delta n_{eff} \equiv 0$) occurs at two points for each value of $E_F$ as we increase $d_1$. More importantly, we see that we can reduce $\Delta n_{eff}$ by up to $0.1$, which is more than enough to compensate for the effective index difference shown in Fig. 3 over the full range of Fermi energies, up to incident fields of at least $15$ kV/cm. 
\begin{figure}[h]
  \centering
  \includegraphics[width=0.8 \columnwidth]{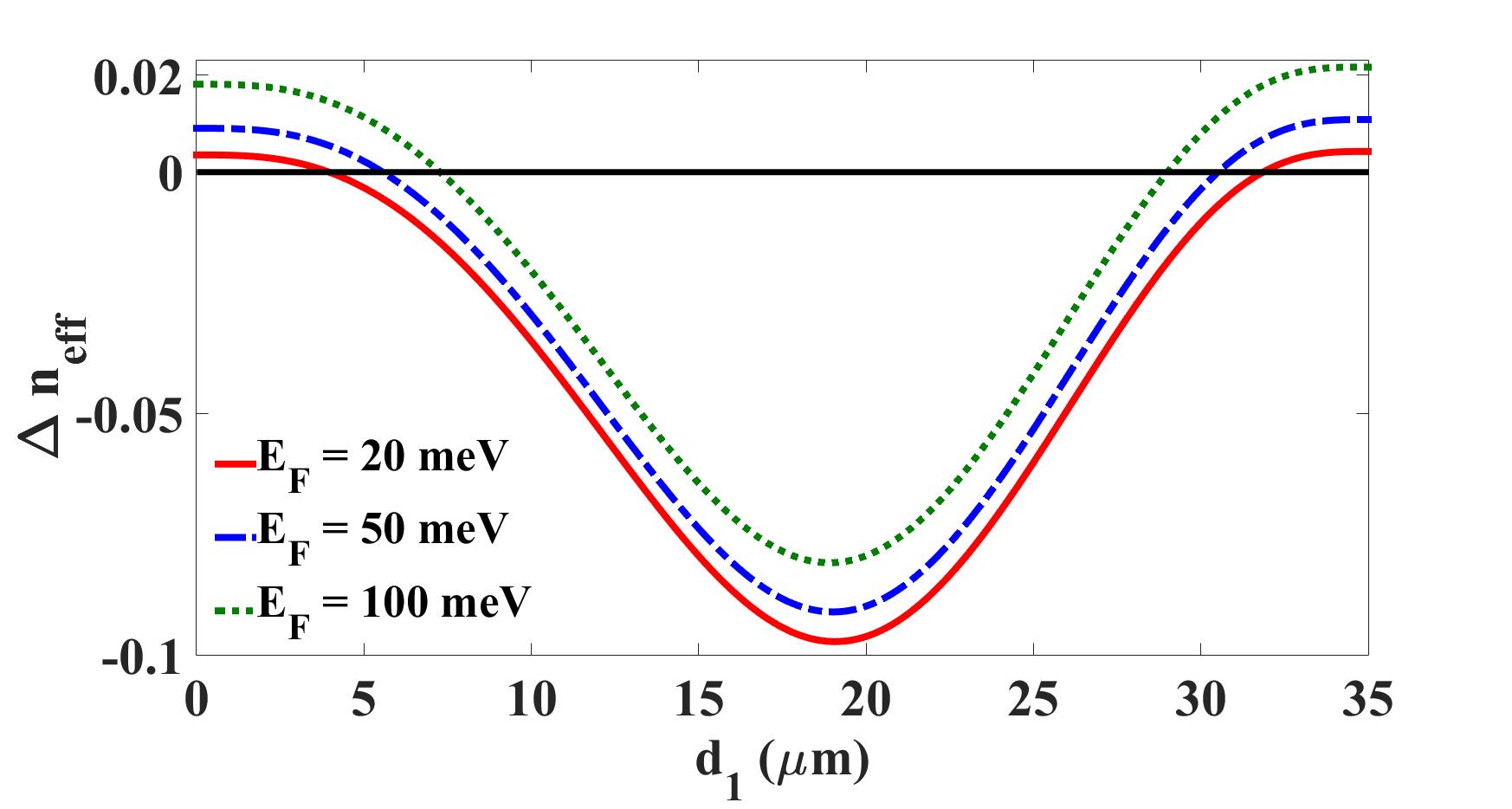} 
  \caption{Effective refractive index difference ( $\Delta n_{eff}$ ) between the TE$_3$ mode at $3\omega$ and the TE$_1$ mode at $\omega$ as a function of $d_1$ for three different Fermi energies. This is calculated without XFM and SFM, i.e. we set $n_2^{(1)} (\omega) = 0$ and $n_2^{(3)}(3\omega) = 0 $ in Eqs.~(21) and (22).}
  \label{fig:Figure8}
\end{figure}
\subsection{Power Efficiency}
In Fig.~8 we plot the power efficiency of the waveguide in our new configuration as a function of $d_1$ for an incident field of $5$ kV/cm at Fermi energies of $E_F = 20$ meV, ${E_F = 50}$~meV, and $E_F = 100$ meV. Decreasing the Fermi energy leads to higher power efficiency due to the increase in the nonlinear conductivity \cite{horng} and reduced loss due to a decrease in the linear conductivity \cite{Alnaib2015,cheng}. Note that when $d_1 = 35 \, \mu m$, our new configuration PPW is identical to our original PPW, and so we obtain the same efficiency as is given in Fig.~4(a). At this low input power, we know from Fig.~3 that the effects of XFM and SFM are almost negligible except for $E_F = 20$ meV. Therefore, we expect to obtain the peaks in the efficiency close to the values of $d_1$ where  $\Delta n_{eff}$ is zero in Fig.~7. This is indeed what we find, but with small shifts that arise from the need to also compensate for XFM and SFM. As expected from Fig.~3, this shift is largest for the Fermi energy of $20$ meV. We also note that for this small input field, the increase in the efficiency over the initial-configuration PPW is rather modest ($< 30\%$).  
\begin{figure}[h]
  \centering
  \includegraphics[width=0.7 \columnwidth]{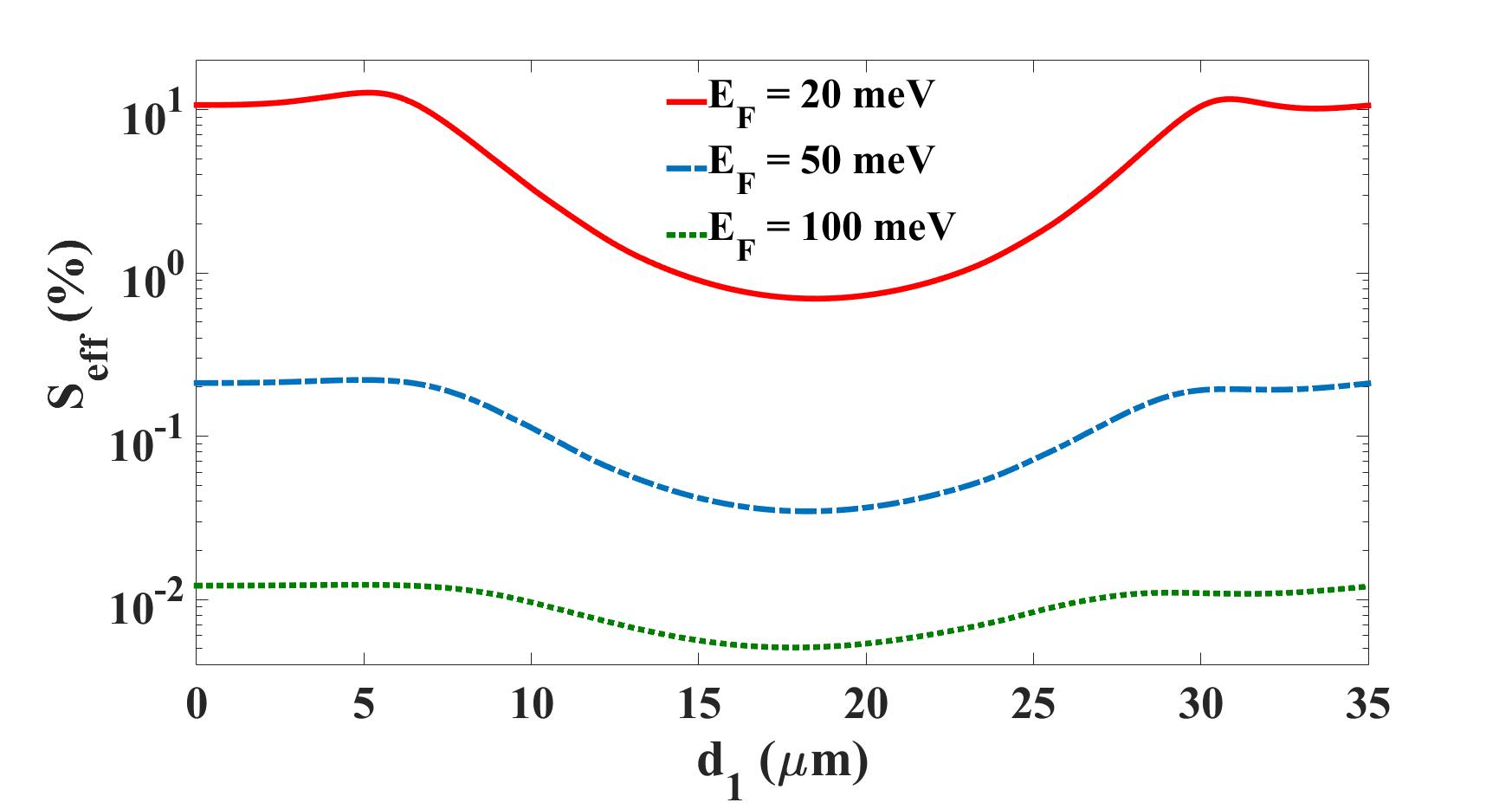} 
  \caption{Power efficiency of the new configuration as a function of $d_1$ for the full calculation for three different Fermi energies and an input field of $E_{input} = 5$ kV/cm.}
  \label{fig:Figure9}
\end{figure}\\
In Fig.~9 we plot the power efficiency of the waveguide in the new configuration as a function of $d_1$ for a Fermi energy of $20$ meV for input fields of $5, 10,$ and $ 15$ kV/cm. As in Fig.~8, the efficiency peaks at two different values for each input field. However, for the higher input fields, we see that these peaks are much larger and have shifted to values of $d_1$ that are closer to the minimum in $\Delta n_{eff}$ seen in Fig.~7. Both of these effects are the result of the need to compensate for the much larger phase mismatch that arises from XFM and SFM at high input fields.
\begin{figure}[h]
  \centering
  \includegraphics[width=0.8 \columnwidth]{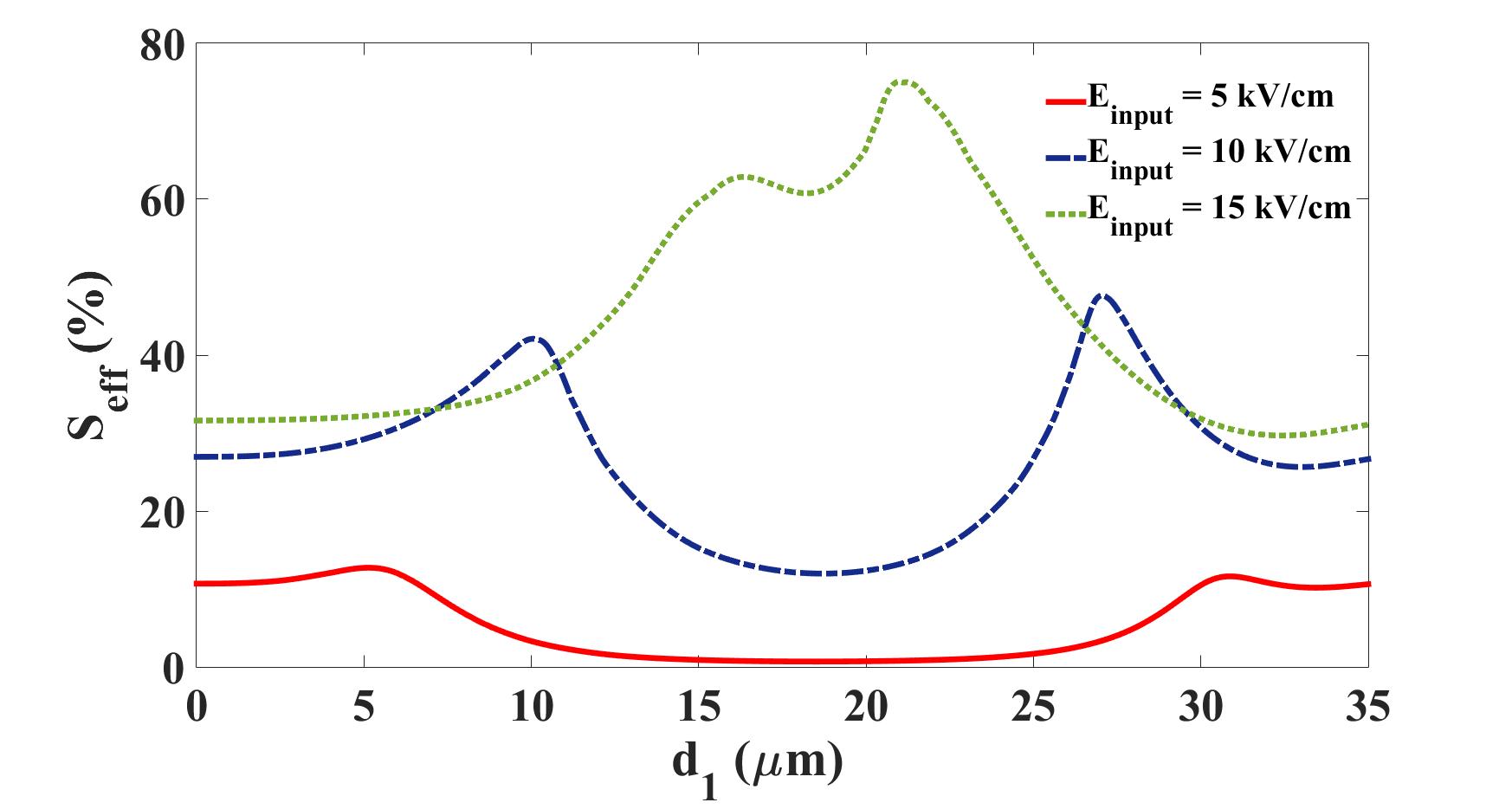} 
  \caption{Power efficiency in the new configuration as a function of $d_1$ for the full calculation for a Fermi energy of $E_F =~20$~meV for three different input fields.}
  \label{fig:Figure10}
\end{figure}\\
In Fig.~10, we compare the power efficiency of the waveguide in the original configuration with and without SFM and XFM to the efficiency of the waveguide in the new configuration with the full calculation. In the new configuration, the optimized power efficiency is improved such that it equals or improves upon the results we obtained for the original configuration when SFM and XFM is neglected. This is because our new configuration  not only overcomes the phase mismatch induced by SFM and XFM but also that induced by the linear response of the graphene. For example, for $E_F = 20$ meV, and $E_{input} = 10$ kV/cm the power efficiency increases from $28\%$ to
$48\%$ and for $E_{input} = 15$ kV/cm it increases
from $35\%$ to $75\%$. 
\begin{figure}[h]
  \centering
  \includegraphics[width=0.7 \columnwidth]{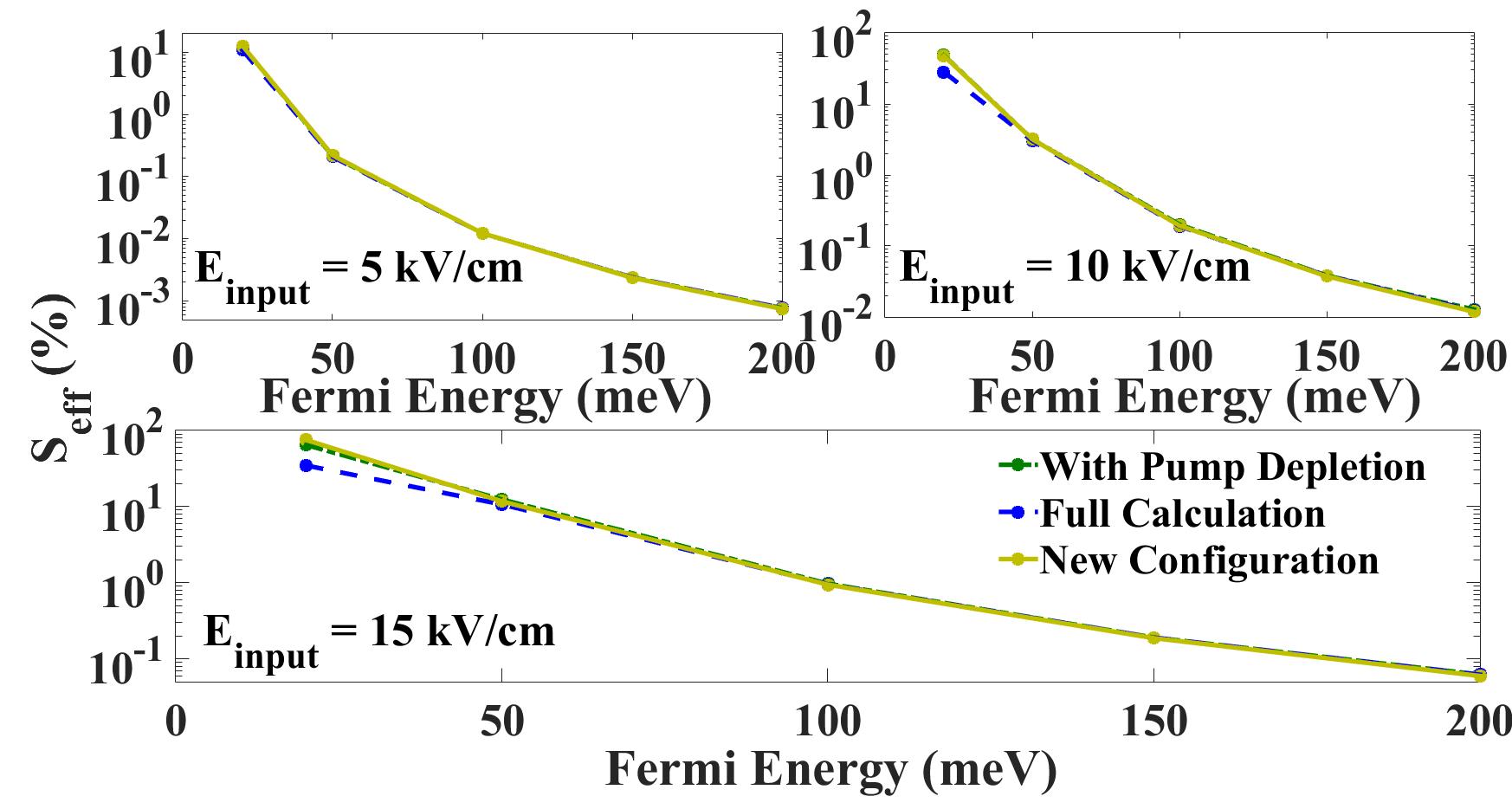} 
  \caption{Power efficiency as a function of Fermi energy for three different input fields of (a) $5$, (b) $10$, and (c) $15$ kV/cm. The dashed curves are for the original PPW with (blue) and without (green) XFM and XFM. The solid curve is the result found including XFM and SFM for optimized new configuration PPW.}
  \label{fig:Figure11}
\end{figure}\\\
Due to the experimental difficulties in achieving a uniform doping over graphene sheets that are millimetres in length, achieving low Fermi energies is very challenging. Therefore, we now examine what efficiencies can be achieved for higher Fermi energies if we move to higher input fields. In Fig.~11(a), we plot the optimized power efficiency of the waveguide in the new configuration as a function of input field for different Fermi energies. Note that the optimized length and $d_1$ are different for each input field and Fermi energy (See Fig.~11(b) and (c)). We note that for all three Fermi energies, the efficiency initially rises with input power, but it then reaches a peak and settles to a value of around $30\%$ at high input fields. To understand this, consider Fig.~3, where we see that for $E_F = 20$ meV, the index mismatch due to XFM and SFM is already $0.06$ at an input field of $15$ kV/cm. It is easy to see therefore that for the fields considered in Fig.~10, we will quickly reach index differences greater than $0.1$ which is the maximum that can be compensated for using our new configuration PPW. Therefore, the input field at which the efficiency peaks for a given Fermi energy is the field at which the nonlinear effective index difference reaches about $0.1$. Due to the strong dependence of XFM and SFM on the Fermi energy, this field amplitude is different for the different Fermi energies. The highest efficiencies are obtained for the lowest Fermi energy of $20$ meV, largely because the loss is lower in this system and because the higher nonlinearity means that the structure length is less. Note that for all three Fermi energies, the peak efficiencies occur for devices with a length of only a few hundred microns. The reason that the high-field efficiency is essentially independent of Fermi energy is because in all cases, the pump is depleted over a distance that is less than the linear loss distance ($1.2$ mm, $0.48$ mm and $0.24$ mm for Fermi energies of $20$ meV, $50$ meV, and $100$ meV, respectively) and so the different losses at the different Fermi energies do not play a significant role. Therefore, we find that very good efficiencies can be obtained for higher Fermi energies if one can attain the higher input fields. For example, for a Fermi energy of $100$ meV, we are able to obtain a power efficiency of $30 \%$ at an input field of $ 50$ kV/cm. Because at high-fields, our structure is not able to compensate for SFM and XFM, we find (not shown) that efficiencies of up to $40\%$ can be obtained at high fields and large Fermi energies even in our original configuration. This is a very promising configuration that we believe should be achievable in the lab using current graphene samples and THz sources \cite{Hafez,Hadi}. 
\begin{figure}[h]
  \centering
  \includegraphics[width=0.9 \columnwidth]{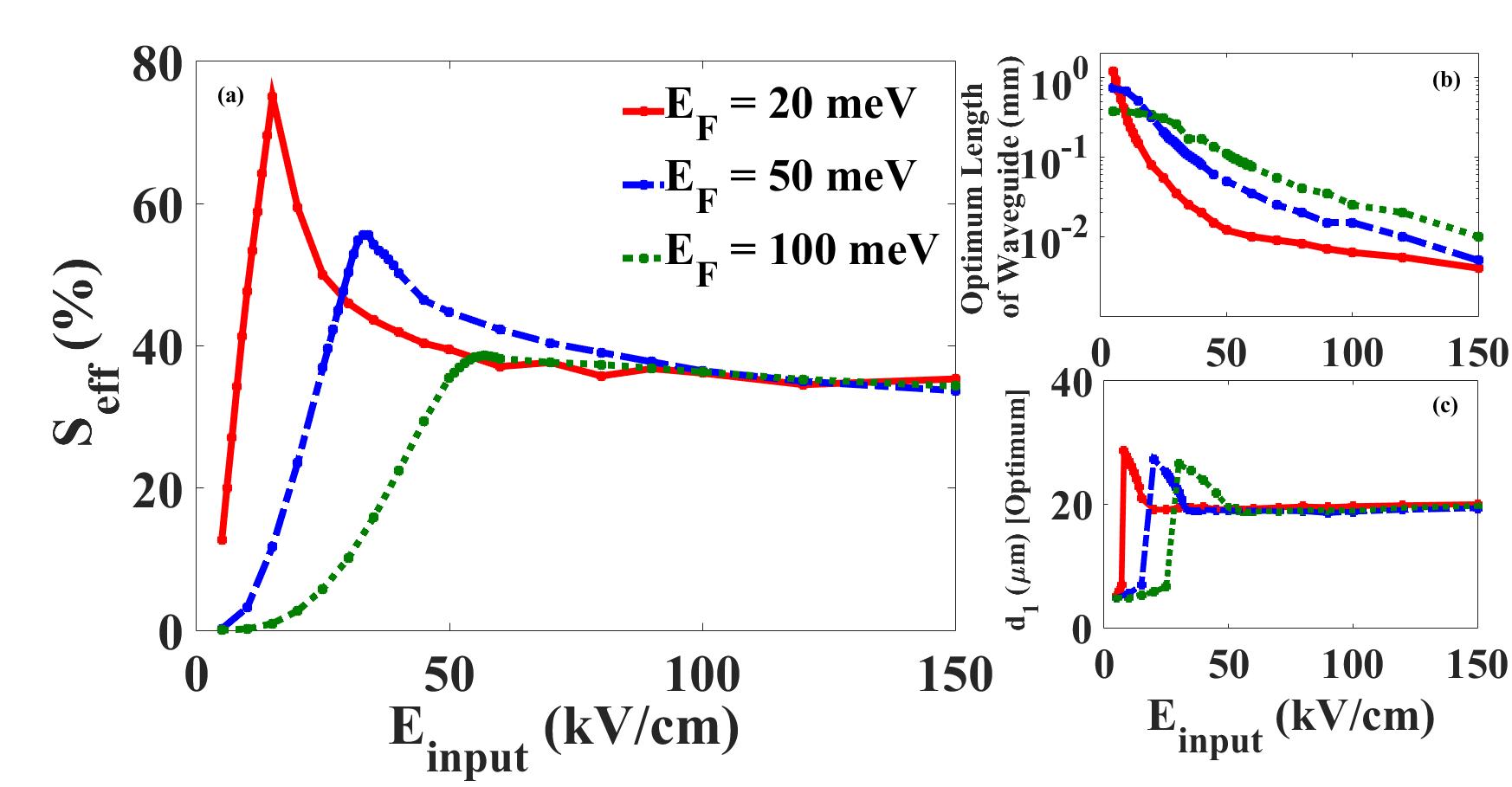} 
  \caption{(a): Power efficiency of the waveguide in the new configuration as a function of input field for Fermi energies of $E_F = 20$, $50$, and $100$ meV. (b) and (c): Optimized length and $d_1$ as a function of input field. }
  \label{fig:Figure11}
\end{figure}
\section{Summary}
We have developed a coupled-mode theory for the propagating lossy modes of the pump and third harmonic fields in a PPW to calculate third harmonic generation, including pump depletion, SFM, and XFM. We find that SFM and XFM degrades the phase matching between the TE$_1$ mode at $\omega$ and TE$_3$ mode at $3\omega$ and thereby decreases the generated third-harmonic electric field. We have shown that one can overcome the phase mismatch due to SFM and XFM by designing a new configuration PPW. We found that by optimizing the dielectric layer thickness, the power efficiency can be increased by more than a factor of two relative to the original configuration. We have also shown that even for graphene with Fermi energy of $100$ meV, where the nonlinearity is relatively modest, efficiencies of up to $30\%$ can be achieved for input field amplitudes of $50$ kV/cm. We therefore believe that our PPW system is an excellent platform to produce and examine harmonic generation in graphene. \\ 
\textbf{Acknowledgements} We thank the Natural Sciences
and Engineering Research Council of Canada and Queen's University for financial support. The authors would like to thank Lukas Helt for useful discussions.
\appendix
\numberwithin{equation}{section}
\numberwithin{figure}{section}
\section{Dynamic equations of the linear and third-harmonic electric field}
In this Appendix, we give the details of the derivation of our coupled-mode equations: Eqs.~(18) and (19). \\
The LHS of Eq.~(17) can be written as 
\begin{equation}
\nabla^2 {E} (y,z;3\omega) = \dfrac{d^2 {E}(y,z;3\omega) }{dy^2} + \dfrac{d^2 {E}(y,z;3\omega)}{dz^2}
\end{equation} 
If we define $S_n^{(1)}\Big(y\Big)  = sin(\tilde{k}_n(\omega) y )$,
\begin{equation}
\dfrac{d^2 {E}(y,z;3\omega) }{dy^2} = \sum_{n} A_n(z,3\omega) e^{i\tilde{\beta}_n(3\omega)z} S_n^{\prime \prime(1)}(y) 
\end{equation}
where
\begin{equation}
S_n^{\prime \prime (1)}\Big(y\Big)  = -\tilde{k}_n^2(\omega) S_n^{(1)}\Big(y\Big).
\end{equation}
However, Eq.~(A.3) is not valid at $y = b/2$. According to the results obtained for the first derivative of the electric field at the graphene, shown in Fig.~A.1, we must add a term to our calculations for the electric field at the graphene. 
\begin{figure}[h]
  \centering
  \includegraphics[width=1 \columnwidth]{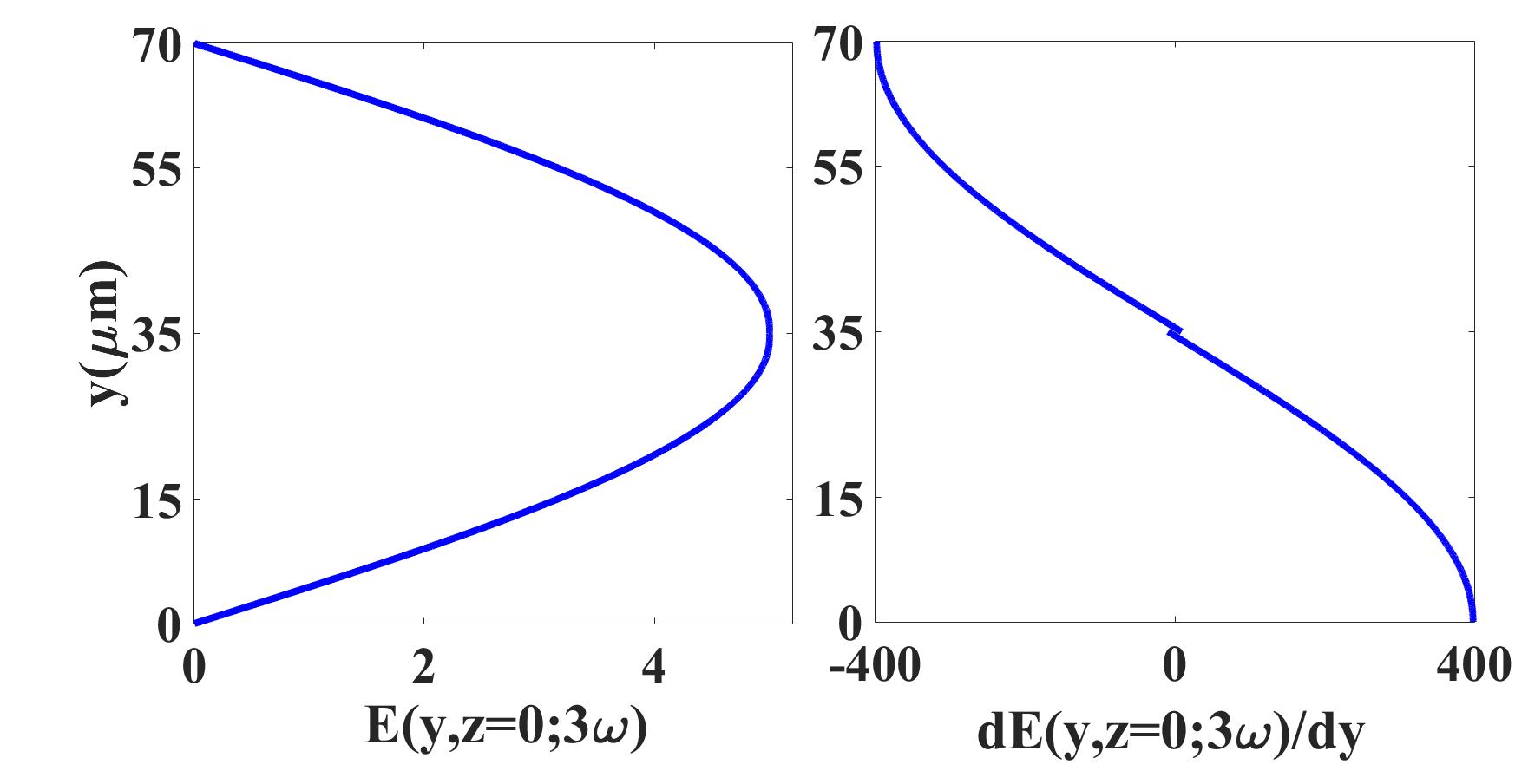} 
  \caption{(a) Electric field and (b) first derivative of the electric field for TE$_1$ mode for a $2$ THz field in the waveguide in the presence of graphene for a plate separation of $b = 70 \, \mu m$. The discontinuity in the mode profile is due to the surface current at the graphene.} 
  \label{fig:Figure1}
\end{figure}
This term is given by $ C \delta (y - b/2)$, 
such that, 
\begin{equation}
S_n^{\prime \prime (1)} \Big(y\Big) = - \tilde{k}_n^2(3\omega) S_n^{(1)}\Big(y\Big) + C \delta(y-b/2).
\end{equation} 
To determine $C$, we take integrate $S_n^{\prime \prime(1)}\Big(y\Big) $ from $y = b/2 - \epsilon$ to $y = b/2 + \epsilon$, for $\epsilon \ll b$. Then, we obtain
\begin{equation}
C = \int_{b/2 -\epsilon}^{b/2 + \epsilon} S_n^{\prime \prime(1)} \Big(y\Big) dy = S_n^{\prime(1)} \Big(y = \dfrac{b}{2}^{+}\Big) - S_n^{\prime(1)} \Big(y = \dfrac{b}{2}^{-}\Big).
\end{equation}\\
The electric field below and above the graphene is defined as
\begin{align}
\mathbf{E}_{below}^{(n)} (y,z;3\omega) =& {E_n} e^{i \tilde{\beta}_n (3\omega)z} \sin(\tilde{k}_n(3\omega) y) \widehat{x}  \quad \quad \quad \quad 0 \;  < \; y \;  < \; b/2  \\
\mathbf{E}_{above}^{(n)}(y,z;3\omega) =& {E_n} e^{i \tilde{\beta}_n (3\omega)z} \sin(\tilde{k}_n(3\omega) (y-b)) \widehat{x}  \quad \quad b/2 \; < \; y  \;  <  \; b \nonumber 
\end{align}
using Maxwell's equations we have,
\begin{align}
\mathbf{\nabla} \times \mathbf{E}_{below}^{(n)} (y,z;3\omega) =& - (-i 3\omega) \mu_0 \mathbf{H}_{below}^{(n)} (y,z;3\omega) \\
\mathbf{\nabla} \times \mathbf{E}_{above}^{(n)} (y,z;3\omega) =& - (-i3 \omega) \mu_0 \mathbf{H}_{above}^{(n)} (y,z;3\omega) \nonumber
\end{align}
Thus, Eq.~(A.7) can be written as 
\begin{align}
\mathbf{H}_{below}^{(n)} (y,z;3\omega) =& \dfrac{\tilde{\beta}_n (3\omega)}{3\omega \mu_0} {E_n} e^{i \tilde{\beta}_n (3\omega)z} \sin(\tilde{k}_n y) \widehat{y} \\
&+ \dfrac{i}{3\omega \mu_0} {E_n} e^{i \tilde{\beta}_n(3\omega) z} \dfrac{d \sin(\tilde{k}_n y)}{dy} \widehat{z} \nonumber, \\
\mathbf{H}_{above}^{(n)} (y,z;3\omega) =& -\dfrac{\tilde{\beta}_n (3\omega)}{3\omega \mu_0}{E_n} e^{i \tilde{\beta}_n (3\omega)z} \sin(\tilde{k}_n (y-b)) \widehat{y} \nonumber \\
&- \dfrac{i}{3\omega \mu_0} {E_n} e^{i \tilde{\beta}_n(3\omega) z} \dfrac{d \sin(\tilde{k}_n (y-b))}{dy} \widehat{z}. \nonumber
\end{align}
The current density at the graphene is given by 
\begin{equation}
\mathbf{J}_L (z; 3\omega) = \sum_n \mathbf{J}_{Ln} (z; 3\omega),
\end{equation}
where the current density at the graphene for $n^{th}$ mode is related to the field by
\begin{align}
\mathbf{J}_{Ln}(z;3\omega) =& 
 -\dfrac{i}{3\omega \mu_0} {E_n} e^{i \tilde{\beta}_n(3\omega)z} \dfrac{d\sin(\tilde{k}_n (y-b))}{dy} \widehat{x} \vert_{y= \dfrac{b}{2}^{+}} \nonumber \\
& -\dfrac{i}{3\omega \mu_0} {E_n} e^{i \tilde{\beta}_n(3\omega)z} \dfrac{d\sin(\tilde{k}_n y)}{dy} \widehat{x} \vert_{y= \dfrac{b}{2}^{-}}, 
\end{align}
where 
\begin{equation}
\dfrac{d \sin(\tilde{k}_n (y-b))}{dy} \vert_{y= \dfrac{b}{2}^{+}}  = -\dfrac{d \sin(\tilde{k}_n y)}{dy} \vert_{y= \dfrac{b}{2}^{-}}. 
\end{equation}
Thus, Eq.~(A.10) can be written as  
\begin{equation}
\mathbf{J}_{Ln}(z;3\omega) = \dfrac{-i}{3\omega \mu_0} {E_n} e^{i \tilde{\beta}_n (3\omega) z} \bigg \lbrace S_n^{\prime (1) } \Big(y= \dfrac{b}{2} ^-\Big) - S_n^{\prime (1)}\Big(y = \dfrac{b}{2}^+\Big) \bigg \rbrace \widehat{x}.
\end{equation}
The current density at the graphene is related to the linear conductivity by
\begin{align}
\mathbf{J}_L (z;3\omega)=& \sum_n \mathbf{J}_{Ln} (z; 3\omega) \\
=& \sigma^{(1)} (3\omega) \mathbf{E}_n(z; 3\omega), \nonumber
\end{align}
where 
\begin{equation}
\mathbf{E}_n(z;3\omega) = {E}_n e^{i \tilde{\beta}_n (3\omega) z} sin(\tilde{k}_n \dfrac{b}{2})\widehat{x}
\end{equation}
Equality of Eq.~(A.12) and Eq.~(A.13) gives
\begin{equation}
S_n^{\prime (1)} \Big(y = \dfrac{b}{2}^{-}\Big) - S_n^{\prime (1)} \Big(y = \dfrac{b}{2}^{+}\Big) = 3i \omega \mu_0 \sigma^{(1)} (3\omega) S_n^{(1)}\Big(\dfrac{b}{2}\Big).
\end{equation}
Using Eq.~(A.15) in Eq.~(A.5), we obtain
\begin{equation}
C = 3 i \omega \mu_0 \sigma^{(1)} (\omega)S_n^{(1)}\Big(\dfrac{b}{2}\Big). 
\end{equation}
Thus, Eq.~(A.4) becomes 
\begin{equation} 
S_n^{\prime \prime (1)} \Big(y\Big) = - \tilde{k}_n^2 S_n^{(1)} \Big(y\Big) - 3i\omega \mu_0 \sigma^{(1)} (3\omega) S_n^{(1)}\Big(\dfrac{b}{2}\Big), 
\end{equation}
and Eq.~(A.2) can be written as
\begin{align}
\dfrac{d^2 {E}(z,y;3\omega)}{d y^2} =& - \sum_{n} \lbrace \tilde{k}_n^2 A_n (z;3\omega) e^{i\tilde{\beta}_n (3\omega)z} S^{(1)}_n \Big(y\Big) \\
&  + 3i\omega \mu_0 \sigma^{(1)} (3\omega) A_n(z;3\omega) S_n^{(1)}\Big(y=\dfrac{b}{2}\Big) \rbrace. \nonumber 
\end{align}
We also have that, using the slowly-varying envelop approximation, where we neglect the second derivative of the envelope, that
\begin{align}
\dfrac{d^2 {E}(z,y;3\omega)}{d z^2 } = \sum_{n} \lbrace & 2i \tilde{\beta}_n (3\omega) \dfrac{d A_n  (z;3\omega)}{dz} \\
&-  \tilde{\beta}_n^2 (3\omega) A_n (z;3\omega) \rbrace e^{i\tilde{\beta}_n (3\omega) z} S_n^{(1)} (y). \nonumber
\end{align}
If we use Eqs.~(A.18) and (A.19) in Eq.~(17) of the main text and retain only those terms that are maximally phase marched, then we obtain
\begin{align}
-&\sum_{n} 2i\tilde{\beta}_n(3\omega) \dfrac{d A_n (z;3\omega)}{d z} e^{i\tilde{\beta}_n(3\omega)z} S_n^{(3)} \Big(y\Big) \\
+& \sum_{n} \lbrace \tilde{\beta}^2_n(3\omega) + \tilde{k}_n^2(3\omega) - 9\mu_0 \epsilon \omega^2 \rbrace A_n (z;3\omega) e^{i\tilde{\beta}_n(3\omega)z} S_n^{(3)} \Big(y\Big) \nonumber\\
+&\sum_{n} 3i\omega \mu_0 \sigma^{(1)} (3\omega) A_n (z;3\omega)e^{i\tilde{\beta}_n(3\omega)z}  S_n^{(3)} \Big(y= b/2\Big)\delta(y-b/2) \nonumber \\
=&  \sum_{n} 3i\omega \mu_0 \sigma^{(1)}(3\omega) A_n(z;3\omega) e^{i\tilde{\beta}_n (3\omega)z} S_n^{(3)} \Big(y\Big) \delta(y-b/2) \nonumber \\
+& 3 i \omega \mu_0 \sigma^{(3)} (3\omega; \omega,\omega,\omega) \sum_{n^{\prime}} \Big\lbrace A_{n^{\prime}} (z;\omega) \Big\rbrace^3 e ^{3i\tilde{{\beta}}_{n^{\prime}}(\omega)z} \Big\lbrace S_{n^{\prime}}^{(1)} \Big(y\Big) \Big\rbrace^3  \delta(y-b/2) \nonumber \\
+& 3 i \omega \mu_0 \sigma^{(3)} (3\omega; 3\omega,-3\omega,3\omega) \sum_{n}  A_{n} (z;3\omega) \vert A_{n} (z;3\omega) \vert^2 e ^{i(2\tilde{{\beta}}_{n}(3\omega) - \tilde{{\beta}}_{n}^{\ast}(3\omega))z} S_{n}^{\ast (3)} \Big(y\Big) \Big\lbrace S_{n}^{(3)} \Big(y\Big) \Big\rbrace^2  \delta(y-b/2) \nonumber \\
+& 6 i \omega \mu_0 \sigma^{(3)} (3\omega; 3\omega,-\omega,\omega) \sum_{n} \sum_{n^{\prime}} A_{n} (z;3\omega) \vert A_{n^{\prime}} (z;\omega) \vert^2 e ^{i(\tilde{{\beta}}_{n}(3\omega)+\tilde{{\beta}}_{n^{\prime}}(\omega) - \tilde{{\beta}}_{n}^{\ast}(\omega))z} S_{n}^{(3)} \Big(y\Big) \vert S_{n^{\prime}}^{(1)} \Big(y\Big) \vert ^2  \delta(y-b/2) \nonumber 
\end{align} 
For a \emph{lossy waveguide}, the propagation constant is defined as $\tilde{\beta}_n (3\omega) = \sqrt{9\mu_0\epsilon\omega^2- \tilde{k}_n^2(3\omega)}$. Using this removes the second term in Eq.~(A.20). We now multiply Eq.~(A.20) by  $S_m^{\ast (1)} \Big(y\Big)$ and integrate over $y$, using  
\begin{align}
\int_0^b {S_m^{\ast}}^{(1)}\Big(y\Big)S_n^{(1)}  \Big(y\Big) dy  \simeq b/2& \delta_{n,m} \\
\int_0^b {S_m^{\ast}}^{(1)} \Big(y\Big) S_n^{(1)} \Big(y\Big) \delta(y-b/2)dy  =&  {S_m^{\ast}}^{(1)}\Big(\dfrac{b}{2}\Big) S_n^{(1)}\Big(\dfrac{b}{2}\Big) \nonumber\\
\int_0^b {S_m^{\ast}}^{(1)}\Big(y\Big) S_{n^{\prime}}^{(3)} \Big(y\Big) \Big\lbrace S_n^{\ast (1)} \Big(y\Big) \Big\rbrace^2 \delta(y-b/2) dy &= {S_m^{\ast}}^{(1)}\Big(\dfrac{b}{2}\Big) S_{n^{\prime}}^{(3)}\Big(\dfrac{b}{2}\Big) \Big\lbrace S_n^{\ast (1)}\Big(\dfrac{b}{2}\Big) \Big\rbrace^2 \nonumber
\end{align}
From this we obtain Eq. (18) for the differential equation of the amplitude of the electric field at $3\omega$ for $m$th mode. A similar calculation yields Eq. (19) for the differential equation for the electric field at $\omega$. Note that we have confirmed numerically that the slowly-varying envelope approximation, where we neglect the second derivatives of the envelopes is an excellent approximation for all fields and Fermi energies.

\end{document}